\documentstyle[12pt]{article}
\newcommand{\beq}{\begin{equation}}
\newcommand{\eeq}{\end{equation}}
\newcommand{\beqs}{\begin{eqnarray}}
\newcommand{\eeqs}{\end{eqnarray}}

\newcommand{\tr}{{\rm Tr}}
\begin{document}
\pagestyle{plain}
\setcounter{page}{1}
\newcounter{bean}
\baselineskip16pt


\begin{titlepage}
\begin{flushright}
PUPT-1716\\
hep-th/9708005
\end{flushright}

\vspace{7 mm}

\begin{center}
{\huge Absorption by Branes and Schwinger Terms }

\vspace{2mm}
{\huge in the World Volume Theory}
\end{center}
\vspace{10 mm}
\begin{center}
{\large  
Steven S.~Gubser and Igor R.~Klebanov \\
}
\vspace{3mm}
Joseph Henry Laboratories\\
Princeton University\\
Princeton, New Jersey 08544
\end{center}
\vspace{7mm}
\begin{center}
{\large Abstract}
\end{center}
\noindent
 We study how coincident Dirichlet 3-branes absorb incident gravitons
polarized along their world volume.  We show that the absorption
cross-section is determined by the central term in the correlator of
two stress-energy tensors. The existence of a non-renormalization
theorem for this central charge in four-dimensional 
${\cal N}=4$ supersymmetric Yang-Mills theories
shows that the leading term at low energies
in the absorption cross-section is not renormalized. This guarantees
that the agreement of the cross-section with semiclassical
supergravity, found in earlier work, survives all loop corrections.
The connection between absorption of gravitons polarized along the
brane and Schwinger terms in the stress-energy correlators of the
world volume theory holds in general. We explore this connection to
deduce some properties of the stress-energy tensor OPE's for 2-branes
and 5-branes in 11 dimensions, as well as for 5-branes in 10
dimensions.

\vspace{7mm}
\begin{flushleft}
July 1997

\end{flushleft}
\end{titlepage}


\newpage
\renewcommand{\baselinestretch}{1.1} 


\renewcommand{\epsilon}{\varepsilon}
\def\equno#1{(\ref{#1})}
\def\equnos#1{(#1)}
\def\sectno#1{section~\ref{#1}}
\def\figno#1{Fig.~(\ref{#1})}
\def\D#1#2{{\partial #1 \over \partial #2}}
\def\df#1#2{{\displaystyle{#1 \over #2}}}
\def\tf#1#2{{\textstyle{#1 \over #2}}}
\def\d{{\rm d}}
\def\e{{\rm e}}
\def\i{{\rm i}}
\def\Leff{L_{\rm eff}}


\def \td {\tilde }
\def \ci {\cite}
\def \sm {$\s$-model }

\def \o {\omega}
\def \inv {^{-1}}
\def \ov {\over }
\def \four{{\textstyle{1\over 4}}}
\def \fourth{{{1\over 4}}}
\def \ha {{1\ov 2}}
\def \QQ {{\cal Q}}


\def\TL{\hfil$\displaystyle{##}$}
\def\TR{$\displaystyle{{}##}$\hfil}
\def\TC{\hfil$\displaystyle{##}$\hfil}
\def\TT{\hbox{##}}

\def\litem#1#2{{\leftskip=20pt\noindent\hskip-10pt {\bf #1} #2\par}}

\def\comment#1{}
\def\fixit#1{}

\def\tf#1#2{{\textstyle{#1 \over #2}}}
\def\df#1#2{{\displaystyle{#1 \over #2}}}

\def\coth{\mathop{\rm coth}\nolimits}
\def\csch{\mathop{\rm csch}\nolimits}
\def\sech{\mathop{\rm sech}\nolimits}
\def\Vol{\mathop{\rm Vol}\nolimits}
\def\vol{\mathop{\rm vol}\nolimits}
\def\diag{\mathop{\rm diag}\nolimits}
\def\tr{\mathop{\rm tr}\nolimits}
\def\Disc{\mathop{\rm Disc}\nolimits}
\def\sgn{\mathop{\rm sgn}\nolimits}

\def\lsim{\mathrel{\mathstrut\smash{\ooalign{\raise2.5pt\hbox{$<$}\cr\lower2.5pt\hbox{$\sim$}}}}}
\def\gsim{\mathrel{\mathstrut\smash{\ooalign{\raise2.5pt\hbox{$>$}\cr\lower2.5pt\hbox{$\sim$}}}}}

\def\slashed#1{\ooalign{\hfil\hfil/\hfil\cr $#1$}}

\def\sqr#1#2{{\vcenter{\vbox{\hrule height.#2pt
         \hbox{\vrule width.#2pt height#1pt \kern#1pt
            \vrule width.#2pt}
         \hrule height.#2pt}}}}
\def\square{\mathop{\mathchoice\sqr56\sqr56\sqr{3.75}4\sqr34}\nolimits}

\def\idget{$\sqr55$\hskip-0.5pt}
\def\endrow{\hskip0.5pt\cr\noalign{\vskip-1.5pt}}
\def\endyoung{\hskip0.5pt\cr}

\def\eff{{\rm eff}}
\def\abs{{\rm abs}}
\def\hc{{\rm h.c.}}
\def\+{^\dagger}
\def\omicron{o}
\def\O{{\cal O}}
\def\cl{{\rm cl}}

\def\M{\cal M}
\def\D#1#2{{\partial #1 \over \partial #2}}

\def\overleftrightarrow#1{\vbox{\ialign{##\crcr
     $\leftrightarrow$\crcr\noalign{\kern-0pt\nointerlineskip}
     $\hfil\displaystyle{#1}\hfil$\crcr}}}
\def\em{\it}   

\def\str{{\rm str}}

\def \t {\tau}
\def \td {\tilde }
\def \ci {\cite}
\def \sm {$\s$-model }

\def \o {\omega}
\def \inv {^{-1}}
\def \ov {\over }
\def \four{{\textstyle{1\over 4}}}
\def \fourth{{{1\over 4}}}
\def \ha {{1\ov 2}}
\def \QQ {{\cal Q}}

\def \lr { \lref}
\def\np {{  Nucl. Phys. }}
\def \pl {{  Phys. Lett. }}
\def \mpl {{ Mod. Phys. Lett. }}
\def \prl {{  Phys. Rev. Lett. }}
\def \pr  {{ Phys. Rev. }}
\def \ap  {{ Ann. Phys. }}
\def \cmp {{ Commun.Math.Phys. }}
\def \ijmp {{ Int. J. Mod. Phys. }}
\def \jmp {{ J. Math. Phys.}}
\def \cqg {{ Class. Quant. Grav. }}

\section{Introduction}
\label{Intro}

Extremal black holes with non-vanishing horizon area may be embedded
into string theory or M-theory using intersecting $p$-branes
\cite{CTT,sv,cm,mst,myers,at,KT,bl}.  These configurations are useful
for a microscopic interpretation of the Bekenstein-Hawking entropy.
The dependence of the entropy on the charges, the non-extremality
parameter, and the angular momentum suggests a connection with $1+1$
dimensional conformal field theory \cite{sv,cm,cy,mst,myers,KT}.  
This ``effective string'' is essentially the intersection
of the $p$-branes.  Calculations of emission and absorption rates
\cite{cm,dmw,dm,GK,mast,GKtwo,CGKT,KM,hawk,dkt,KK,KRT,Mathur,Gubser,CL}
provide further tests of the ``effective string'' models of $D=5$
black holes with three charges and of $D=4$ black holes with four
charges.  For minimally coupled scalars the functional dependence of
the greybody factors on the frequency agrees exactly with
semiclassical gravity, providing a highly non-trivial verification of
the effective string idea \cite{mast,GKtwo}. Similar successes have
been achieved for certain non-minimally coupled scalars, which were
shown to couple to higher dimension operators on the effective string.
For instance, the ``fixed'' scalars \cite{Kallosh} were shown to
couple to operators of dimension $(2,2)$ \cite{CGKT} while the
``intermediate'' scalars \cite{KRT} -- to operators of dimension
$(2,1)$ and $(1,2)$. Unfortunately, there is little understanding of
the ``effective string'' from first principles, and some of the more
sensitive tests reveal this deficiency. For instance, semiclassical
gravity calculations of the ``fixed'' scalar absorption rates for
general black hole charges reveal a gap in our understanding of higher
dimension operators \cite{KK}.  A similar problem occurs when one
attempts a detailed effective string interpretation of the higher
partial waves of a minimally coupled scalar \cite{Mathur,Gubser}. Even
the s-wave absorption by black holes with general charges is complex
enough that it is not reproduced by the simplest effective string
model \cite{KM,CL}. These difficulties by no means invalidate the
general qualitative picture, but they do pose some interesting
challenges.  In order to gain insight into the relation between
gravity and D-branes it is useful to study, in addition to the
intersecting branes, the simpler configurations which involve parallel
branes only.

A microscopic interpretation of the entropy of near-extremal
$p$-branes was first studied in \cite{GKP,kt}. It was found that the
scaling of the Bekenstein-Hawking entropy with the temperature agrees
with that for a massless gas in $p$ dimensions only for the
``non-dilatonic $p$-branes'': namely, the self-dual 3-brane of the type
IIB theory, and the 2- and 5-branes of M-theory.  Their further study
was undertaken is \cite{IK,GKT} where low-energy absorption
cross-sections for certain incident massless particles were compared
between semiclassical supergravity and string or M-theory. For the
3-branes, exact agreement was found for the leading low-energy
behavior of the absorption cross-sections for dilatons \cite{IK}, as
well as for R-R scalars and for gravitons polarized along the brane
\cite{GKT}.  The string-theoretic description of macroscopic 3-branes
can be given in terms of many coincident D3-branes \cite{dlp,polch}.
Indeed, $N$ parallel D3-branes are known to be described by a $U(N)$
gauge theory in 3+1 dimensions with ${\cal N}=4$ supersymmetry
\cite{EW}. This theory has a number of remarkable properties,
including exact S-duality, and we will be able to draw on a 
known non-renormalization theorem in explaining the absorption by the
3-branes.

 From the point of view of supergravity, the 3-brane is also
special because its extremal geometry,
\beq
d s^2 =\left (1+{R^4\over r^4}\right )^{-1/2}
(-dt^2 + dx_1^2 + dx_2^2+ dx_3^2 ) +
\left (1+ {R^4\over r^4}\right )^{1/2} (dr^2 + r^2 d\Omega_5^2)\ ,
\eeq
is non-singular \cite{ght},
while the dilaton background is constant.  Instead of a singularity 
at $r=0$ we
find an infinitely long throat whose radius is determined by the
charge (the vanishing of the horizon area is due to the longitudinal
contraction).  Thus, for a large number $N$ of coincident branes, the
curvature may be made arbitrarily small in Planck units. For instance,
for $N$ D3-branes, the curvature is bounded by a quantity of order
 \beq
{1\over \sqrt{N\kappa}} \sim {1\over \alpha'\sqrt{N g_{\rm str}}}
 \ .
\eeq
In order to suppress the string scale corrections to the classical
metric, we need to take the limit $N g_{\rm str}\rightarrow\infty $.

The tension of a D3-brane depends on $g_{\rm str}$ and $\alpha'$ only
through the ten-dimensional gravitational constant $\kappa = 8
\pi^{7/2} g \alpha'^2$:
 \beq
T_{(3)}={\sqrt \pi \over\kappa}\ . 
\eeq
 This suggests that we can compare the expansions of various
quantities in powers of $\kappa$ between the microscopic and the
semiclassical descriptions. Indeed, the dominant term in the
absorption cross-section at low energy is \cite{IK,GKT}
 \beq
\label{three}
\sigma= {\pi^4\over 8}\omega^3 R^8
={\kappa^2  \omega^3 N^2\over 32 \pi} \ , 
\eeq
which agrees between semiclassical supergravity and string theory.

It is important to examine the structure of higher power in $g_{\rm
str}$ corrections to the cross-section \cite{Das}.  In semiclassical
supergravity the only quantity present is $\kappa$, and corrections to
(\ref{three}) can only be of the form
 \beq\label{ACor}
  a_1 \kappa^3 \omega^7+ a_2 \kappa^4 \omega^{11}+ \ldots
 \eeq
 However, in string theory we could in principle find corrections even
to the leading term $\sim \omega^3$, so that
 \beq\label{BCor}
\sigma_{\rm string}= {\kappa^2  \omega^3 N^2\over 32 \pi} 
\left (1+ b_1 g_{\rm str} N+ b_2 (g_{\rm str} N)^2+ \ldots \right )
+ {\cal O} (\kappa^3 \omega^7)\ .\eeq
 Presence of such corrections would spell a manifest disagreement with
supergravity because, as we have explained, the comparison has to be
carried out in the limit $N g_{\rm str}\rightarrow\infty $.  The
purpose of this paper is to show that, in fact, $b_i=0$ due to a
non-renormalization theorem in $D=4$ ${\cal N}=4$ SYM theory.

In section 2 we present a detailed argument for the absence of such 
corrections in the absorption cross-section of gravitons polarized
along the brane. These gravitons couple to the stress-energy tensor
on the world volume and we will show that the absorption cross-section is,
up to normalization, the central term in the two-point function of
the stress-energy tensor. The fact that $b_i=0$ follows from the fact
that the one-loop calculation of the central charge is exact
in $D=4$ ${\cal N}=4$ SYM theory.

The connection between absorption of gravitons polarized along the
brane and Schwinger terms in the stress-energy correlators of
the world volume theory is
a general phenomenon that holds for all branes. 
In section 3 we explore this connection to deduce some properties
of the stress-energy tensor OPE's for multiple 2-branes and 5-branes
of M-theory, as well as for multiple 5-branes of string theory.

\section{D-brane approach to absorption}

It was probably Callan who first realized that, in terms of D-brane
models, absorption cross-sections correspond up to a simple overall
factor to discontinuities of two point functions of certain operators
on the D-brane world volume \cite{cp}.  This realization was exploited
in \cite{ja,GSpin}.  Consider massless scalar particles in ten
dimensions normally incident upon D3-branes.  If the coupling to the
brane is given by 
  \begin{equation}\label{GenCoup}
   S_{\rm int} = \int d^4 x \, \phi(x,0) {\cal O}(x) \ ,
  \end{equation}
 where $\phi(x,0)$ is a canonically normalized scalar field evaluated
on the brane, and ${\cal O}$ is a local operator on the brane, then
the precise correspondence is
  \begin{equation}\label{SigmaDisc}
   \sigma = {1 \over 2 i \omega} \Disc \Pi(p) 
    \bigg|_{p^0 = \omega \atop \vec{p} = 0} \ .
  \end{equation}
 Here $\omega$ is the energy of the incident particle, and
  \begin{equation}\label{PiP}
   \Pi(p) = \int d^4 x \, e^{i p \cdot x} 
    \langle {\cal O}(x) {\cal O}(0) \rangle \ .
  \end{equation}
 When ${\cal O}$ is a scalar in the world volume theory, $\Pi(p)$
depends only on $s = p^2$, and $\Disc \Pi(p)$ is computed as the
difference of $\Pi$ evaluated for $s = \omega^2 + i \epsilon$ and $s =
\omega^2 - i \epsilon$.  In the case of the graviton, we shall see
that $\Pi(p)$ is a polynomial in $p$ times a function of $s$, so the
evaluation of $\Disc \Pi(p)$ is equally straightforward.

The validity of (\ref{SigmaDisc}) depends on $\phi$ being a
canonically normalized field.  In form it is almost identical to the
standard expression for the decay rate of an unstable particle of mass
$\omega$.  The dimensions are different, however: $\sigma$ is the
cross-section of the 3-brane per unit world volume, and so has
dimensions of $(\hbox{length})^5$.  Similar formulas can be worked out
for branes of other dimensions and for near-extremal branes, although
away from extremality (\ref{PiP}) would become a thermal Green's
function (see \cite{ja,GSpin}).

Now let us work through the example of the graviton.  The 3-brane
world volume theory is ${\cal N}=4$ supersymmetric $U(N)$ gauge
theory, where $N$ is the number of parallel 3-branes \cite{EW}.
Thus, the massless fields on the world volume are the gauge field, six
scalars, and four Majorana fermions, all in the adjoint representation
of $U(N)$.  To lowest order in $\kappa$, and ignoring the couplings to
the bulk fields, the world volume-action is ($I=1,...,4$; $i=4,...,9$)
  \beq
   S_3 = \int d^4 x \, \tr \left[ -\tf{1}{4} F_{\alpha\beta}^2 + 
    \tf{i}{2} \bar\psi^I \gamma^\alpha \partial_\alpha \psi_I + 
    \tf{1}{2} (\partial_\alpha X^i)^2 + 
     \hbox{interactions} \right] \ .  \label{wvAction}
  \eeq
 The interactions referred to here are the standard renormalizable
ones of ${\cal N} = 4$ super-Yang-Mills (see for example
\cite{Sohnius} for the complete flat-space action).  In equation (3.4)
of \cite{GKT}, a factor of the 3-brane tension $T_{(3)}$ appeared
in front of the action.  It is convenient to work with canonically
normalized fields, and so, relative to the conventions of \cite{GKT},
we have absorbed a factor of $\sqrt{T_{(3)}}$ into $A_\mu$, $\psi^I$,
and $X^i$.  Another difference between \cite{GKT} and the present
paper is that we work here with a mostly minus metric and the spinor
conventions of \cite{Sohnius}.

The full (and as yet unknown) action for multiple 3-branes is
non-polynomial, and (\ref{wvAction}) includes only the dimension $4$
terms.  The higher dimension terms will appear with powers of
$1/T_{(3)}$, which is to say positive powers of $\kappa$.  They could
give rise to corrections of the form (\ref{ACor}), but not
(\ref{BCor}).

Despite our ignorance of the full action for multiple coincident
3-branes, one can be fairly confident in asserting that the
external gravitons polarized parallel to the brane couple via 
 \beq \label{sint}
    S_{\rm int} = \int d^4 x \, \tf{1}{2} h^{\alpha\beta} 
     T_{\alpha\beta}  \ ,
 \eeq 
 where $h_{\alpha\beta} = g_{\alpha\beta} - \eta_{\alpha\beta}$ is the
perturbation in the metric, and $T_{\alpha\beta}$ is the stress-energy
tensor:
  \beq\label{ConfT}
   \vcenter{\openup1\jot
   \halign{\strut\span\TL & \span\TR\cr
    T_{\alpha\beta} &= \tr \big[ -F_{\alpha}^{\ \gamma} F_{\beta\gamma} + 
      \tf{1}{4} \eta_{\alpha\beta} F_{\gamma\delta}^2 +
     \tf{i}{2} \bar\psi^I \gamma_{(\alpha} \partial_{\beta)} \psi_I  
      \cr
     &\quad + 
      \tf{2}{3} \partial_\alpha X^i \partial_\beta X^i  - 
       \tf{1}{6} \eta_{\alpha\beta} (\partial_\gamma X^i)^2 -
       \tf{1}{3} X^i \partial_\alpha \partial_\beta X^i + 
       \tf{1}{12} \eta_{\alpha\beta} X^i \square X^i  \cr
     &\quad +
      \hbox{interactions} \, \big] \ .  \cr
   }}
  \eeq 
 This ``new improved'' form
of the stress-energy tensor \cite{ccj} is chosen so that $\partial_\mu
T^{\mu\nu} = 0$ and $T^\mu_\mu = 0$ on shell.  The scalar terms differ
from the canonical form
  \beq
   T^{\rm (can)}_{\alpha\beta} = 
    \tr \left[ \partial_\alpha X^i \partial_\beta X^i - 
     \tf{1}{2} \eta_{\alpha\beta} (\partial_\gamma X^i)^2 + 
     \ldots \, \right] \ .
  \eeq
 The difference arises from adding a term $-\tf{1}{12} \sqrt{-g} R \tr
X^2$ to the lagrangian so that the scalars are conformally coupled.

Choosing the scalars to be minimally or conformally coupled does not
affect the one-loop result for the cross-section.  But it is the
traceless form of $T_{\alpha\beta}$ presented in (\ref{ConfT}) which
has a non-renormalized two-point function.  
The reason is that the conformal
$T_{\alpha\beta}$ is in the same supersymmetry multiplet as the
supercurrents and the $SU(4)$ R-currents.  ${\cal N} = 4$
super-Yang-Mills theory is finite to all orders and anomaly free in
flat space.  The Adler-Bardeen theorem guarantees that any anomalies
of the $SU(4)$ R-currents can be computed exactly at one loop.
Because $\partial_\alpha R^\alpha$ and $T^\alpha_\alpha$ are in the
same supermultiplet (the so-called ``multiplet of anomalies''), the
one-loop result for the trace anomaly must also be exact.\footnote{We
thank D. Anselmi for pointing out the relevance of the Adler-Bardeen
theorem.}  Thus the trace anomaly in a curved background is the same
as in the free theory:
  \beq\label{TraceAnom}
   \langle T^\mu_\mu \rangle = 
    -{1 \over 16 \pi^2} \left( c F - 
      {2 c \over 3} \square R - b G \right) \ ,
  \eeq
 where $F = C_{\alpha\beta\gamma\delta}^2$ is the square of the Weyl
tensor and $G = R_{\alpha\beta\gamma\delta}^2 - 4 R_{\alpha\beta}^2 +
R^2$ is the topological Euler density.  Thus the second and third
terms in (\ref{TraceAnom}) are total derivatives.  The coefficients
$c$ and $b$ are given by \cite{BD}
  \beq\label{CentralC}
   \vcenter{\openup1\jot
   \halign{\strut\span\TL & \span\TR\cr
    c &= {12 N_1 + 3 N_{1/2} + N_0 \over 120} = {N^2 \over 4}  \cr
    b &= {124 N_1 + 11 N_{1/2} + 2 N_0 \over 720} = {N^2 \over 4} \ . \cr
   }}
  \eeq
 $N_1 = N^2$, $N_{1/2} = 4 N^2$, and $N_0 = 6 N^2$ are the numbers of
spin-one, Majorana spin-half, and real spin-zero fields in the
super-Yang-Mills theory.  Note that spin-one, spin-half, and spin-zero
particles make contributions to $c$ in the ratio $2:2:1$.  The
different spins contribute to the cross-section in precisely the same
ratio, as was demonstrated in \cite{GKT}.

It remains to make the connection between $\langle T^\alpha_\alpha
\rangle$ and $\langle T_{\alpha\beta}(x) T_{\gamma\delta}(0) \rangle$.
Suppressing numerical factors and Lorentz structure, the OPE of
$T_{\alpha\beta}$ with $T_{\gamma\delta}$ is \cite{AnsOne}
  \beq\label{TOPE}
   T(x) T(0) = {c \over x^8} + \ldots + {T(0) \over x^4} + \ldots \ .
  \eeq
 We have omitted terms involving the Konishi current as well as terms
less singular than $1/x^4$, and we have anticipated the conclusion
that the coefficient on the Schwinger term is precisely the central
charge $c$ appearing in (\ref{TraceAnom}).  A clean argument to this
effect is presented in \cite{oe} and summarized briefly below.  The
Schwinger term is nothing but the two-point function: with all factors
and indices written out explicitly\footnote{The field normalization
conventions in \cite{AnsTwo} differ from those used here.}
\cite{AnsTwo},
  \beq\label{TPF}
   \langle T_{\alpha\beta}(x) T_{\gamma\delta}(0) \rangle = 
    {c \over 48 \pi^4} X_{\alpha\beta\gamma\delta} 
    \left( 1 \over x^4 \right)
  \eeq
 where
  \beq\label{XDef}
   \vcenter{\openup1\jot
   \halign{\strut\span\TL & \span\TR\cr
     X_{\alpha\beta\gamma\delta} &= 
      2 \square^2 \eta_{\alpha\beta} \eta_{\gamma\delta} - 
       3 \square^2 (\eta_{\alpha\gamma} \eta_{\beta\delta} + 
        \eta_{\alpha\delta} \eta_{\beta\gamma}) -
      4 \partial_\alpha \partial_\beta 
        \partial_\gamma \partial_\delta  \cr
      &\quad - 2 \square 
        (\partial_\alpha \partial_\beta \eta_{\gamma\delta} +
         \partial_\alpha \partial_\gamma \eta_{\beta\delta} +
         \partial_\alpha \partial_\delta \eta_{\beta\gamma} +
         \partial_\beta \partial_\gamma \eta_{\alpha\delta} +
         \partial_\beta \partial_\delta \eta_{\alpha\gamma} +
         \partial_\gamma \partial_\delta \eta_{\alpha\beta}) \ .  \cr
   }}
  \eeq
 The argument of \cite{oe} starts by obtaining an expression for the
flat space three point function $\langle T^\alpha_\alpha(x)
T_{\beta\gamma}(y) T_{\rho\sigma}(z) \rangle$ by expanding
(\ref{TraceAnom}) around flat space to second order.  Using Ward
identities for the conservation of $T_{\mu\nu}$, one can then derive a
relation on two point functions which can only be satisfied if the
coefficients $c$ in (\ref{TPF}) and (\ref{TraceAnom}) are identical.

We are finally ready to compute the cross-section.  Fourier
transforming the two-point function (\ref{TPF}), one obtains
  \beq\label{PiPT}
   \Pi_{\alpha\beta\gamma\delta}(p) = \int d^4 x \, 
    e^{i p \cdot x} 
    \langle T_{\alpha\beta}(x) T_{\gamma\delta}(0) \rangle = 
    {c \over 48 \pi^4} \hat{X}_{\alpha\beta\gamma\delta} 
     \int d^4 x \, {e^{i p \cdot x} \over x^4}
  \eeq
 where $\hat{X}_{\alpha\beta\gamma\delta}$ is just the
$X_{\alpha\beta\gamma\delta}$ of (\ref{XDef}) with $\partial \to -i p$.
The integral is evaluated formally as 
  \beq\label{FormalInt}
   \Pi(s) = \int d^4 x \, {e^{i p \cdot x} \over x^4} = 
    \pi^2 \log(-s) + \hbox{(analytic in s)}
  \eeq
 where, as before, $s = p^2$.  One easily reads off the discontinuity
across the positive real axis in the $s$-plane: 
  \beq
   \Disc \Pi(s) = \Pi(s + i \epsilon) - \Pi(s - i \epsilon) = 
    -2 \pi^3 i \ ,
  \eeq
 and so
  \beq
   \Disc \Pi_{\alpha\beta\gamma\delta}(p) = -{i c \over 24 \pi} 
    \hat{X}_{\alpha\beta\gamma\delta} \ .
  \eeq
 For the sake of definiteness, let us consider a graviton polarized in
the $x^1$--$x^2$ direction.  $\hat{X}_{1212} = -3 \omega^4$ for
normal incidence, so
  \beq\label{SigmaGrav}
   \sigma = {2 \kappa^2 \over 2 i \omega} \Disc \Pi_{1212}(p)
     \bigg|_{p^0 = \omega \atop \vec{p} = 0}
    = {c \over 8 \pi} \kappa^2 \omega^3 \ ,
  \eeq
 in agreement with the classical result when $c = N^2 / 4$.  The extra
factor of $2 \kappa^2$ in the second expression in (\ref{SigmaGrav})
comes from the fact that $h_{\alpha\beta}$ as defined in the text
following (\ref{sint}) is not a canonically normalized scalar field;
instead, $h_{\alpha\beta} / \sqrt{2 \kappa^2}$ is.

In summary, the non-renormalization argument is as follows: the
graviton cross-section is read off at leading order in $\kappa$, but
correct to all orders in $g_{\rm str}$, from the two-point function of
the stress-energy tensor.  The two-point function is not renormalized
beyond one loop because the Schwinger term in the OPE (\ref{TOPE}) is
similarly non-renormalized.  That in turn is due to the fact that the
central charge appearing in the Schwinger term is precisely the
coefficient of the Weyl tensor squared in the trace anomaly
(\ref{TraceAnom}).  The trace anomaly is not renormalized past one
loop because $T^\alpha_\alpha$ is related by supersymmetry to the
divergence of the $SU(4)$ $R$-current, which is protected by the
Adler-Bardeen theorem against anomalies beyond one loop.  Another,
more heuristic, reason why the central charge should not be
renormalized is that there is a critical line extending from $g_{\rm
str} N=0$ to $g_{\rm str}N= \infty$.  The central charge is expected
to be constant along a critical line.  In four dimensions, this
expectation is supported by the work of \cite{AnsTwo}.\footnote{A more
definitive argument has been given in four dimensions for the
constancy of flavor central charges along fixed lines
\cite{AnsThree}.}  Therefore, $c$ can be calculated in the $g_{\rm
str} N\rightarrow 0$ limit where it is given by one-loop diagrams.

We could adopt a different strategy and invert our arguments.
Requiring that the world volume theory of $N$ coincident 3-branes
agrees with semiclassical supergravity tells us that in the $g_{\rm
str} N\rightarrow \infty$ limit its central charge approaches
$N^2/4$. In the next section we will similarly deduce the Schwinger
terms in the two-point functions of the stress-energy tensor for other
branes. Furthermore, we can study two-point functions of other
operators by calculating the semiclassical absorption cross-section
for particles that couple to them. For instance, the dilaton couples
to $\tr F^2$. The dilaton absorption cross-section was calculated in
\cite{IK}. The semiclassical result implies that the Schwinger term
here is again $\sim N^2$ in the $g_{\rm str} N\rightarrow \infty$
limit.  Comparison with the gauge theory calculation \cite{IK,GKT}
suggests that the one-loop result is again exact. It will be
interesting to extract more results from these connections between
gravity and gauge theory.

\section{Extension to Other Branes}

In this section we further explore the connection between the
absorption cross-sections in semiclassical supergravity and central
terms in two-point functions calculated in corresponding world
volume theories. We proceed in analogy to the 3-brane discussion
presented in the previous section, and consider absorption of
gravitons polarized along the branes. In the world volume theories
such gravitons couple to components of stress-energy tensor,
$T_{\alpha\beta}$. On the other hand, in supergravity such
gravitons satisfy the minimally coupled scalar equation with
respect to the coordinates transverse to the brane \cite{GKT}.
This establishes a general connection between the absorption
cross-section of a minimally coupled scalar and the central
term in the algebra of stress-energy tensors.
For $N$ coincident 3-branes the world volume theory
is known, and we have shown that the conformal anomaly is
in exact agreement with this principle. There are cases, however,
where little is known about the world volume theory of multiple
branes. In such cases we can use our method to find the Schwinger
term without knowing any details of the world volume
theory.

Consider, for instance, the 2-branes and the 5-branes of M-theory.
While the world volume theory of multiple coincident branes
is not known in detail, the extreme
supergravity solutions are well-known. The absorption
cross-sections for low-energy
gravitons polarized along the brane were calculated
in \cite{IK,GKT,emp}, with the results
\beq
\sigma_2= {1\over 6\sqrt 2 \pi} \kappa_{11}^2 \omega^2 N^{3/2}\ ,
\qquad\qquad
\sigma_5={1\over 3\cdot 2^6 \pi^2} N^3\omega^5\ .
\eeq
For $N$ coincident M2-branes we may deduce that the schematic
structure of the stress-energy tensor OPE is
\beq
T(x) T(0) = {c_2\over x^6} + \ldots
\eeq
where the central charge behaves as $c_2\sim N^{3/2}$
in the large $N$ limit.
For $N$ coincident M5-branes we instead have
\beq
T(x) T(0) = {c_5\over x^{12}} + \ldots \ .
\eeq
Now the central charge behaves as $c_5\sim N^3$
in the large $N$ limit.
These results have an obvious connection with properties of
the near-extremal entropy found in \cite{kt}.
Indeed, the near-extremal entropy of a large number $N$
of coincident M2-branes is formally
reproduced by ${\cal O}(N^{3/2})$
massless free fields in 2+1 dimensions, while that of
$N$ coincident M5-branes is reproduced by ${\cal O}(N^3)$
massless free fields in 5+1 dimensions. 

As a final example we consider the 5-branes of string theory.
In \cite{juan} it  was shown that their near-extremal entropy
is reproduced by a novel kind of string theory, rather than by
massless fields in 5+1 dimensions. For $N$ coincident D5-branes
the string tension turns out to be that of a D-string divided by
$N$. This suggests that the degrees of freedom responsible for the
near-extremal entropy are those of ``fractionated'' D-strings bound to
the D5-branes.  An S-dual of this picture suggests that the entropy
of multiple NS-NS 5-branes comes from fractionated fundamental
strings bound to them.\footnote{New insights into
the world volume theory of NS-NS 5-branes were recently
obtained in \cite{dvv,ns,witten}.} We would like to learn more about these
theories by probing them with longitudinally polarized
gravitons incident transversely to the brane.

The extreme Einstein metric of both the NS-NS and the R-R 5-branes
is
\beq
d s_E^2 = \left (1 + {R^2\over r^2}\right )^{-1/4}
(-d t^2+ dx_1^2+ \ldots + dx_5^2)+
\left (1 + {R^2\over r^2}\right )^{3/4} (dr^2 + r^2 d\Omega_3^2)\ .
\eeq
The s-wave Laplace equation in the background of this metric is
\beq
\left [ \rho^{-3} {d\over d\rho}
 \rho^3 {d\over d\rho} + 1 + {(\omega R)^2\over \rho^2}
\right ] \phi (\rho)=0\ ,
\eeq
where $\rho =\omega r$.
Remarkably, this equation is exactly solvable in terms of 
Bessel functions. The two possible solutions are
\beq \rho^{-1} J_{\pm \sqrt{1-(\omega R)^2}} (\rho)
\ .\eeq 
Clearly, there are two physically different regimes.
For $\omega R > 1$ the label of the Bessel function is
imaginary, and the requirement that the wave is incoming for
$\rho\rightarrow 0$ selects the solution
\beq \rho^{-1} J_{- i \sqrt{(\omega R)^2- 1}} (\rho)
\ .\eeq 
{}From the large $\rho$ asymptotics we find that the absorption
probability is
\beq
{\cal P} = 1-
e^{- 2\pi \sqrt{(\omega R)^2- 1}} \ .
\eeq
Hence, for $\omega R\geq 1$ the absorption cross-section is 
\beq \label{cs}
\sigma = {4\pi\over \omega^3} \left (1-
e^{- 2\pi \sqrt{(\omega R)^2- 1}} \right )\ .
\eeq
For $\omega R < 1$ the question of how to choose the solution
is somewhat more subtle. It is clear that
$\rho^{-1} J_{\sqrt{1-(\omega R)^2}} (\rho)$ is better behaved near
$\rho =0$ than
$\rho^{-1} J_{-\sqrt{1-(\omega R)^2}} (\rho)$.
If we approach the extreme 5-brane as a limit of a near-extreme
5-brane, we indeed find that
$\rho^{-1} J_{\sqrt{1-(\omega R)^2}} (\rho)$ is the solution
that is selected. Since this solution is real, there is no
absorption for $\omega R <1$.\footnote{The fact that an
extremal 5-brane in 10 dimensions 
does not absorb minimally coupled scalars
below a certain threshold was noted in \cite{BL}.}
This result agrees with the extremal
limit of the 5-brane absorption cross-section calculated in
\cite{KM}.

It is not hard to generalize our calculation to higher partial waves.
For the $\ell$-th partial wave we find that the absorption
cross-section vanishes for $\omega R \leq \ell+1$. Above this threshold
it is given by
\beq 
\sigma_\ell = {4\pi (\ell+1)^2\over \omega^3} \left (1-
e^{- 2\pi \sqrt{(\omega R)^2- (\ell+ 1)^2}} \right )\ .
\eeq

{}From (\ref{cs}) we reach the surprising conclusion that 
  \beq
   \langle T(\omega, \vec 0) T(-\omega, \vec 0) \rangle
  \eeq
 vanishes identically for $\omega <1/R$, which implies that gravity
does not couple to the massless modes of the world volume theory!  The
threshold energy $1/R$ is precisely $1/\sqrt{\alpha'_{\rm eff}}$,
where $\alpha'_{\rm eff} = 1 / (2 \pi T_{\eff})$ and
  \beq
   T_{\rm eff}= {1\over 2\pi R^2}
  \eeq
 is the tension of the fractionated strings.  The ordinary superstring
has its first massive excited state at mass $m^2 = 2/ \alpha'$.  The
threshold energy squared is half this value with $\alpha'$ replaced
by $\alpha'_{\rm eff}$.  If one imagines producing a single massive
string at $\omega = 1/R$, then its mass is $m = 1/\sqrt{\alpha'_{\rm
eff}}$.  Perhaps this is the first excited level of the non-critical
string living on the 5-brane. Similarly, the higher partial wave
thresholds might correspond to higher excited levels of mass
$(\ell+1)/\sqrt{\alpha'_{\rm eff}}$. If instead $\omega \geq 1/R$
corresponds to the pair production threshold of the first massive
state of fractionated strings, then the mass would be $m = 1 / (2
\sqrt{\alpha'_{\rm eff}})$.  Neither picture yields any obvious
explanation of the behavior
  \beq\label{scalingE}
   \langle T(\omega, \vec 0) T(-\omega, \vec 0) \rangle \sim
   \left[ (\omega R)^2- 1 \right]^{1/2}
  \eeq
 just above threshold.  Pair production in a weakly interacting theory
would predict a $7/2$ power in (\ref{scalingE}).  It would be
interesting to find an explanation of the observed square-root scaling
in (\ref{scalingE}).

We believe that our discussion applies to a large number of coincident
D5-branes, as well as to solitonic 5-branes of type IIA and IIB
theories. This is because all these solutions have the same Einstein
metric. Perhaps in the large $N$ limit some properties of the world
volume theories of these different branes become
identical.\footnote{We thank J. Maldacena for suggesting this
possibility to us.}
For $N$ coincident NS-NS 5-branes,
\beq
R^2\sim N\alpha'\ .
\eeq
Thus, the absorption cross-section (\ref{cs}) is formally independent
of $g_{\rm str}$. Therefore, our formula should
be applicable in the $g_{\rm str}\rightarrow 0$
limit proposed in \cite{ns}. There it was argued that in this limit
the NS-NS 5-branes decouple from the bulk modes. We indeed find that
incident gravitons are not absorbed for sufficiently low energies.
However, above a critical energy of order $1/\sqrt{N\alpha'}$
the 5-branes do appear to couple to the bulk modes. As we have 
commented, this is probably related to the fact that the scale
of the string theory living on the 5-brane is
\beq
\alpha'_{\rm eff} = N\alpha'\ ,
\eeq
i.e. the fundamental strings become fractionated \cite{juan}.

In summary, we note that probing branes with low-energy particles
incident from transverse directions is a useful tool for extracting
correlation functions in their world volume theory.  Here we have
given an application of this technique to M2-branes and to 5-branes,
but a more general investigation would be worthwhile.

\section*{Acknowledgments}

We are grateful to C. Callan, O. Ganor, J. Maldacena, M. Perry,
S. Shenker, W. Taylor, A. Tseytlin, and E. Witten for useful
discussions, and to the Aspen Center for Physics for its hospitality.
This work was supported in part by the DOE grant DE-FG02-91ER40671,
the NSF Presidential Young Investigator Award PHY-9157482, and the
James S.{} McDonnell Foundation grant No.{} 91-48.  S.S.G.{} thanks
the Hertz Foundation for its support.



\end{document}